\documentclass[journal]{IEEEtran}

\ifCLASSINFOpdf
\else
   \usepackage[dvips]{graphicx}
\fi
\usepackage{url}
\usepackage{babel}
\usepackage{amsmath,amsfonts}

\usepackage{graphicx}
\usepackage{xcolor}
\usepackage{algorithm}
\usepackage{algpseudocode}

\newcommand{\ing}[1]{\mathsf{#1}}
\newcommand{\Rn}[1]{\ifthenelse{\equal{#1}{}}{\mathbb{R}}{\mathbb{R}^{\ing{#1}}}}
\newcommand{\Cn}[1]{\mathbb{C}^{\ing{#1}}}
\newcommand{\Rset}[2]{\ifthenelse{\equal{#2}{1}}{\in \Rn{\ing{#1}}}{\in \Rn{\ing{#1} \times \ing{#2}}}}
\newcommand{\Cset}[2]{\ifthenelse{\equal{#2}{1}}{\in \Cn{\ing{#1}}}{\in \Cn{\ing{#1} \times \ing{#2}}}}
\newcommand{\vect}[1]{\boldsymbol{\mathbf{\MakeLowercase{#1}}}}
\newcommand{\mtrx}[1]{\boldsymbol{\mathbf{#1}}}

\newcommand{\transp}[1]{#1^{\mathsf{T}}}

\newcommand{\ind}[2]{\ifthenelse{\equal{#2}{}}{\chi_{#1}}{\chi_{#1}\left( #2 \right)}}

\DeclareMathOperator*{\argmax}{argmax\,}

\begin{document}

\title{Generalized Time Domain Velocity Vector}

\author{Sr\dj{}an Kiti\'c and J\'er\^ome Daniel\\Orange Labs
\thanks{Both authors have equally contributed to the submitted article.}}

\maketitle

\begin{abstract}

We introduce and analyze Generalized Time Domain Velocity Vector (GTVV), an extension of the previously presented acoustic multipath footprint extracted from the Ambisonic recordings. GTVV is better adapted to adverse acoustic conditions, and enables efficient parameter estimation of multiple plane wave components in the recorded multichannel mixture. Experiments on simulated data confirm the predicted theoretical advantages of these new spatio-temporal features.


\end{abstract}


\IEEEpeerreviewmaketitle

\section{Introduction}
\label{sec:intro}

Estimating the direction of an active sound source is an important task for a number of applications, including speech enhancement \cite{xenaki2018sound}, robot navigation \cite{evers2018acoustic}, compression \cite{pulkki2007spatial} or augmented reality \cite{ribeiro2012auditory}. On the one hand, localizing an acoustic source is often made difficult by acoustic reflections and reverberation that interfere with direct wave, which generally causes an angular bias and uncertainty \cite{blandin2012multi,daniel2020time}. On the other hand, the knowledge of directions and delays of reflected waves, with respect to the direct one, enables some innovative use-cases such as the inference of room geometry \cite{dokmanic2013acoustic}, source separation aided by the acoustic echoes \cite{scheibler2018separake}, or localization of a source hidden by a soundproof obstacle \cite{kitic2014hearing}.

In \textit{First Order Ambisonics (FOA)}, \emph{i.e.} the first order spherical harmonic (SH) representation of a sound field \cite{zotter2019ambisonics,fernandez2016compressive}, a common source localization method is through the use of so-called \textit{pseudo-intensity} \cite{jarrett2017theory}, \emph{i.e.} an estimate of \textit{active sound intensity}~\cite{rossing2007springer}. In favorable acoustic conditions (free field propagation, relatively low noise), this vector quantity is orthogonal to the propagating wavefront, and allows for the Direction-of-Arrival (DoA) estimation at low computational cost  \cite{merimaa2006analysis,jarrett2017theory}. Unfortunately, in the reverberant environments, the predictions produced by this method are often biased. 
Some methods to make the intensity estimate more robust aim at identifying the time-frequency bins dominated by the direct sound \cite{moore2015direction}, or on exploiting the higher order spherical harmonic coefficients (\emph{a.k.a.} \textit{Higher Order Ambisonics (HOA)}), such as in \cite{moore2016direction,hafezi2017augmented}. 

While the common wisdom in state-of-the-art was, thus, to suppress the effects of acoustic reflections, in the earlier work \cite{daniel2020time}, we have shown that it is possible to improve the DoA estimate \textit{and} estimate the direction and the delay of a dominant horizontal acoustic reflection, at the same time. To achieve this, we use two analysis tools: \textit{Frequency Domain Velocity Vector (FDVV)}, and \textit{Time Domain Velocity Vector (TDVV)}. The former is a normalized version of complex sound intensity \cite{rossing2007springer,merimaa2006analysis}, aggregating both active intensity (in-phase part), and \textit{reactive} intensity (quadrature part), while the latter is its temporal representation. We have demonstrated that TDVV has an interpretable structure that enables straightforward estimation of the acoustic multipath parameters: the propagating directions, relative delays and attenuation factors. However, the TDVV analysis has been theoretically justified only under a restrictive condition that the magnitude of the direct wave exceeds the cumulative magnitude of the reflections. 

In this article we generalize the TDVV framework in two ways: first, we include in the representation the HOA coefficients of any order, and second, the reference component is allowed to be a linear combination of all channels. Both extensions bring their own benefits: while the addition of HOA improves the discrimination of two (or more) wavefront directions, the appropriate choice of the reference component enables the time domain analysis, even under the adverse acoustic conditions. We remark that, when the reference component is the omnidirectional (zero-order) Ambisonic channel, the frequency and time domain formulations are equivalent to \textit{Relative Transfer Function (ReTF)} \cite{gannot2001signal,jarrett2017theory} (also known as ``relative harmonic coefficients'' in some recent works \cite{hu2020unsupervised,hu2020semi}), and \textit{Relative Impulse Response (ReIR)} \cite{talmon2009relative}, in spherical harmonics domain, respectively. To the best knowledge of the authors, the only other work that mentions non-conventional  reference component is \cite{biderman2016efficient}, which discusses the ReTF/ReIR in spherical harmonic domain for a denoising application, under the assumption that the directions of acoustic reflections are already known.

The article is organised as follows: in Section~\ref{sec:GTDVV} we introduce Generalized Time domain Velocity Vector (GTVV) and discuss its properties. Next, in Section~\ref{sec:estimation} a robust estimator of GTVV is presented. The inference of acoustic parameters from an estimated GTVV is discussed in Section~\ref{sec:inference}. The proof-of-concept experimental results are given in Section~\ref{sec:experiments}. The Section~\ref{sec:conclusion} concludes the article.

\section{Generalized Time domain Velocity Vector}
\label{sec:GTDVV}

Consider a single, far field sound source emitting the signal $S(f)$ at frequency $f$, in an reverberant environment. Our aim is to model the received signal which includes the direct path propagation, as well as several dominant acoustic reflections, which are the delayed and attenuated copies of $S(f)$. 
Assuming that mode strength compensation \cite{jarrett2017theory} has been applied at the recording microphone, the spherical harmonics expansion coefficient $B^{(\ing{lm})}(f)$, of order $\ing{l}$ and degree $\ing{m}$, due to $\ing{N}+1$ incoming acoustic waves (``early'' reverberation), is modelled as follows \cite{jarrett2017theory,zotter2019ambisonics}:
\begin{equation}\label{eqMixtureModel}
    B^{(\ing{lm})}(f) = S(f) \sum\limits_{\ing{n}=0}^{\ing{N}} a_{\ing{n}}(f) Y^{(\ing{lm})}(\Omega_{\ing{n}}).
\end{equation}
The frequency-dependent factor $a_{\ing{n}}(f)$ models the complex amplitude $c_{\ing{n}}(f)$ and phase shift of the $\ing{n}$\textsuperscript{th} wavefront, \emph{i.e.} $a_{\ing{n}}(f) = c_{\ing{n}}(f) e^{-j2 \pi f \bar{\tau}_{\ing{n}}}$, where $\bar{\tau}_{\ing{n}}$ is its Time-of-Arrival (ToA). The scalar $Y^{(\ing{lm})}(\Omega_{\ing{n}})$ represents the (order $\ing{l}$, degree $\ing{m}$) spherical harmonic function evaluated at the wavefront direction ${\Omega_{\ing{n}} := (\theta_{\ing{n}}, \varphi_{\ing{n}})}$, with $\theta_{\ing{n}}$ and $\varphi_{\ing{n}}$ being its azimuth and elevation coordinates, respectively. 

Let $\vect{y}_{\ing{n}}$ designate the vector whose entries are the SH coefficients (up to some maximal order $\ing{l}=\ing{L}$), corresponding to the $\ing{n}$\textsuperscript{th} wavefront only:
\begin{equation}
    \vect{y}_{\ing{n}}:=\vect{y}(\Omega_{\ing{n}})=\transp{[Y^{(00)}(\Omega_{\ing{n}}) \; \transp{\vect{y}^{(1*)}(\Omega_{\ing{n}})} \dots \; \transp{\vect{y}^{(\ing{L}*)}(\Omega_{\ing{n}})}]},
\end{equation}
where each subvector $\vect{y}^{(\ing{l}*)}(\Omega_{\ing{n}})$ is composed of $2\ing{l}+1$ coefficients of order $\ing{l}$, and $\transp{(\cdot)}$ is the transpose. Then the observation vector $\vect{b}(f)$, consisting of stacked expansion coefficients $\left\{ B^{(\ing{lm})}(f) \right\}_{\ing{l}\in[0,\ing{L}],\ing{m}\in[-l,l]}$ can be represented as 
\begin{equation}
    \vect{b}(f) = S(f) \sum\limits_{\ing{n}=0}^{\ing{N}} a_{\ing{n}}(f) \vect{y}_{\ing{n}}.
\end{equation}

We denote by $\vect{v}(f)$ the \textit{Generalized Frequency domain Velocity Vector (GFVV)}, defined as follows:
\begin{equation}\label{eqFDVVinst}
    \vect{v}(f) = \frac{\vect{b}(f)}{\transp{\vect{w}(f)}\vect{b}(f) } = \frac{ \sum\limits_{\ing{n}=0}^{\ing{N}} a_n(f) \vect{y}_{\ing{n}}}{ \sum\limits_{\ing{n}=0}^{\ing{N}} a_{\ing{n}}(f) 
    \beta_{\ing{n}}(f)},
\end{equation}
where $\vect{w}(f)$ is a complex weight vector (\emph{e.g.}, a beamformer), and   $\beta_{\ing{n}}(f)=\transp{\vect{w}(f)}\vect{y}_{\ing{n}}$. In the noiseless setting, provided that $S(f)$ is non-negligible, GFVV does not depend on the content of the source signal. We remark that the frequency domain velocity vector (FDVV), introduced in \cite{daniel2020time}, is a particular case of GFVV, for which $\vect{b}$ is restricted to $4$ FOA channels, and $\vect{w}=\transp{[1 \; \transp{\vect{0}}]}$, \emph{i.e.} the reference component is the omnidirectional channel $B^{(00)}(f)$ ($\vect{0}$ is the all-zero vector of an appropriate size).

Let ${a_{\ing{n}}(f)/a_{\ing{p}}(f) = g_{\ing{n}}(f) e^{-j2\pi f (\bar{\tau}_{\ing{n}} - \bar{\tau}_{\ing{p}})}}$, \emph{i.e.} we declare $g_{\ing{n}}(f) = c_{\ing{n}}(f)/c_{\ing{p}}(f)$, which is a relative gain between two wavefronts. Without loss of generality, let $\ing{p}=0$ and set $\beta_0 = 1$, which gives
\begin{equation}\label{eqFDVV}
    \vect{v}(f) = \frac{\vect{y}_0 + \sum\limits_{\ing{n}=1}^{\ing{N}} \frac{\gamma_{\ing{n}}}{\beta_{\ing{n}}} \vect{y}_{\ing{n}} }{1 + \sum\limits_{\ing{n}=1}^{\ing{N}} \gamma_{\ing{n}} },
\end{equation}
where $\gamma_{\ing{n}} = g_{\ing{n}}(f) \beta_{\ing{n}}(f) e^{-j2\pi f (\bar{\tau}_{\ing{n}} - \bar{\tau}_0)}$ (thus, $\gamma_0=1$).


Assume now that $|\sum_{\ing{n}=1}^{\ing{N}} \gamma_{\ing{n}}(f) | <1$: a sufficient condition for this inequality to hold would be ${\sum_{\ing{n}=1}^{\ing{N}} |g_{\ing{n}}(f) \beta_{\ing{n}}(f)| < 1}$. This is rarely the case for FDVV \cite{daniel2020time} or ReTF \cite{jarrett2017theory,hu2020unsupervised,hu2020semi}, for which the reference component is omnidirectional ($\beta_{\ing{n}} = 1$ for all $\ing{n}$). However, a suitable choice of the filter $\vect{w}(f)$ could ensure that this condition is satisfied. For example, if $\vect{w}$ was a signal-independent beamformer \cite{jarrett2017theory}, it could be oriented towards $\Omega_0$, such that the remaining directions are sufficiently attenuated. If this is indeed the case, the denominator part of the eq.~\eqref{eqFDVV} admits a Taylor (specifically, geometric) series expansion:
\begin{equation}
    \rho := \frac{1}{1 + \sum\limits_{\ing{\ing{n}}=1}^{\ing{N}} \gamma_{\ing{n}}} = \sum\limits_{\ing{k}=0}^{\infty} \left(-\sum\limits_{\ing{n}=1}^{\ing{N}} \gamma_{\ing{n}} \right)^{\ing{k}} = \sum\limits_{\ing{k}=0}^{\infty} \rho_{\ing{k}},
\end{equation}
where
\begin{equation*}
     \rho_{\ing{k}} = \sum_{\ing{i}_1 + \ing{i}_2 + \hdots + \ing{i}_N = k} \frac{\ing{k}!}{\ing{i}_1!\ing{i}_2!\hdots \ing{i}_{\ing{N}}!} \prod\limits_{\ing{q}=1}^{\ing{N}} (\gamma_{\ing{n}})^{\ing{q}}.
\end{equation*}

This expression is cumbersome, however, one may observe that each $\rho_{\ing{k}}$ would have an $\ing{k}$\textsuperscript{th}-order term 
that involves only the parameters of a single wavefront, while the remainder contains ``cross-terms'' between different wavesfronts. These cross-terms are grouped together under the variable $\eta(f)$:
\begin{equation}
    \rho = 1 + \sum\limits_{\ing{k}=1}^{\infty}\sum\limits_{\ing{q}=1}^{\ing{N}} \left( -g_{\ing{q}}(f)\beta_{\ing{q}}(f) \right)^{\ing{k}} e^{-j2\pi f \ing{k} (\bar{\tau}_{\ing{q}} - \bar{\tau}_0)} + \eta(f).
\end{equation}
Since $|g_{\ing{q}}(f)\beta_{\ing{q}}(f)|<1$, the higher order polynomial terms in the expression above have progressively decreasing magnitudes (the same holds true for the $\eta(f)$ part).

At this point we make one further assumption: we consider the gains $g_{\ing{n}}$ to be frequency-independent. In practice, the frequency response of the gains depends on the environment: for instance, it may be dictated by the impedance of the materials that generate acoustic reflections~\cite{rossing2007springer}. Hence, this may not be a fully plausible assumption, but one could argue that, in many practical settings, the impedance varies smoothly within certain frequency range. Importantly, considering $g_{\ing{n}}$ to be a constant reduces the number of unknowns in our model. Although it is a parameter we are in control of, we also choose the frequency-independent spatial filter ($\vect{w}(f) = \vect{w}$, thus $\beta_{\ing{n}}(f)=\beta_{\ing{n}}$ for all frequencies), for mathematical convenience, since it simplifies the ensuing time domain expressions.

Hence, the GFVV is expressed as
\begin{multline}
    \vect{v}(f) = \left( \vect{y}_0 + \sum\limits_{\ing{n} = 1}^{\ing{N}} g_{\ing{n}} \beta_{\ing{n}} e^{-j2\pi f (\bar{\tau}_{\ing{n}} - \bar{\tau}_0)}\vect{y}_{\ing{n}} \right) \\
    \times \left( 1 + \sum\limits_{\ing{k} \geq 1} \sum\limits_{\ing{q}=1}^{\ing{N}} (-1)^{\ing{k}} g_{\ing{q}}^{\ing{k}} \beta_{\ing{q}}^{\ing{k}} e^{-j2\pi f \ing{k}(\bar{\tau}_{\ing{q}} - \bar{\tau}_0)} + \eta(f) \right).
\end{multline} 

Applying the inverse Fourier transform to $\vect{v}(f)$, 
and manipulating the resulting expression, produces GTVV:
\begin{multline}\label{eqGTVV}
    \vect{v}(t) = \delta(t) \vect{y}_0 +  \\
    \sum\limits_{\ing{k}\geq 1}\sum\limits_{\ing{n}=1}^{\ing{N}} (-g_{\ing{n}}\beta_{\ing{n}})^{\ing{k}} \left( \vect{y}_0 - \frac{1}{\beta_{\ing{n}}} \vect{y}_{\ing{n}} \right)\delta(t - \ing{k}(\bar{\tau}_{\ing{n}} - \bar{\tau}_0)) + \tilde{\eta}(t),
\end{multline}
where again the terms depending on multiple wavefronts are isolated in the single variable $\tilde{\eta}(t)$. Note that $\tilde{\eta}(t)$ has discrete support on the temporal axis, which never includes $t\leq0$ and, usually, does not coincide with $\delta(t - \ing{k}(\bar{\tau}_{\ing{n}} - \bar{\tau}_0))$ either. 
The GTVV  structure in \eqref{eqGTVV} reveals very useful properties, telling us that a ``well-behaved'' GTVV is necessarily causal (\emph{i.e.}, ${\vect{v}(t <0 ) = \vect{0}}$), and sparse. And equally important, evaluating GTVV at $t=0$ provides an immediate estimate of $\vect{y}_0$.

\section{Estimation of GFVV and GTVV}
\label{sec:estimation}

The energy of speech signals varies over time and frequency, hence we cannot directly use the expression \eqref{eqFDVVinst} to estimate the GFVV. Particularly, as some frequency bins have negative Signal-to-Noise Ratio (SNR), such an estimate would be prone to errors, or even numerical instabilities. Instead, we adapt a well-known ReTF estimator presented in \cite{gannot2001signal}, that exploits the (non)stationarity properties of speech and noise signals \cite{shalvi1996system}. We chose this approach due to its simplicity and the fact that it does not require a prior knowledge of voice activity, but other estimators could be used as well, \emph{e.g.} based on covariance subtraction or whitening \cite{jarrett2017theory}. 

For reasons that will become obvious, in this section we use a time-frequency representation of the considered quantities, such as Short-Time Fourier Transform (STFT). 
Let $N^{(\ing{lm})}(\ing{u},f)$ denote the additive noise of the $(\ing{l,m})$\textsuperscript{th} channel of $\vect{b}(\ing{u},f)$, assumed independent of the source signal $S(\ing{u},f)$, at frame $\ing{u}$ and frequency $f$. The goal is to estimate a ratio of source signal-excited parts of the given Ambisonic channel and the reference, hence one can write:
\begin{equation}\label{eqRearranged}
    B^{(\ing{lm})}(\ing{u},f) = v^{(\ing{lm})}(f)\sum\limits_{\ing{l}',\ing{m}'} w^{(\ing{l}'\ing{m}')} B^{(\ing{l}'\ing{m}')}(\ing{u},f) + U^{(\ing{lm})}(\ing{u},f),
\end{equation}
with
\begin{equation*}
U^{(\ing{lm})}(\ing{u},f) = N^{(\ing{lm})}(\ing{u},f) - v^{(\ing{lm})}(f)\sum\limits_{\ing{l}',\ing{m}'} w^{(\ing{l}'\ing{m}')} N^{(\ing{l}'\ing{m}')}(\ing{u},f).
\end{equation*}
The variables $v^{(\ing{lm})}(f)$ and $w^{(\ing{lm})}$ denote the $(\ing{l,m})$\textsuperscript{th} entries of $\vect{v}(f)$ and $\vect{w}$, respectively. Note that $v^{(\ing{lm})}(f)$ does not depend on $\ing{u}$: this is valid when neither the source nor the microphone are mobile during the considered temporal segment.

\begin{figure}
    \centering
    \includegraphics[width=0.9\columnwidth]{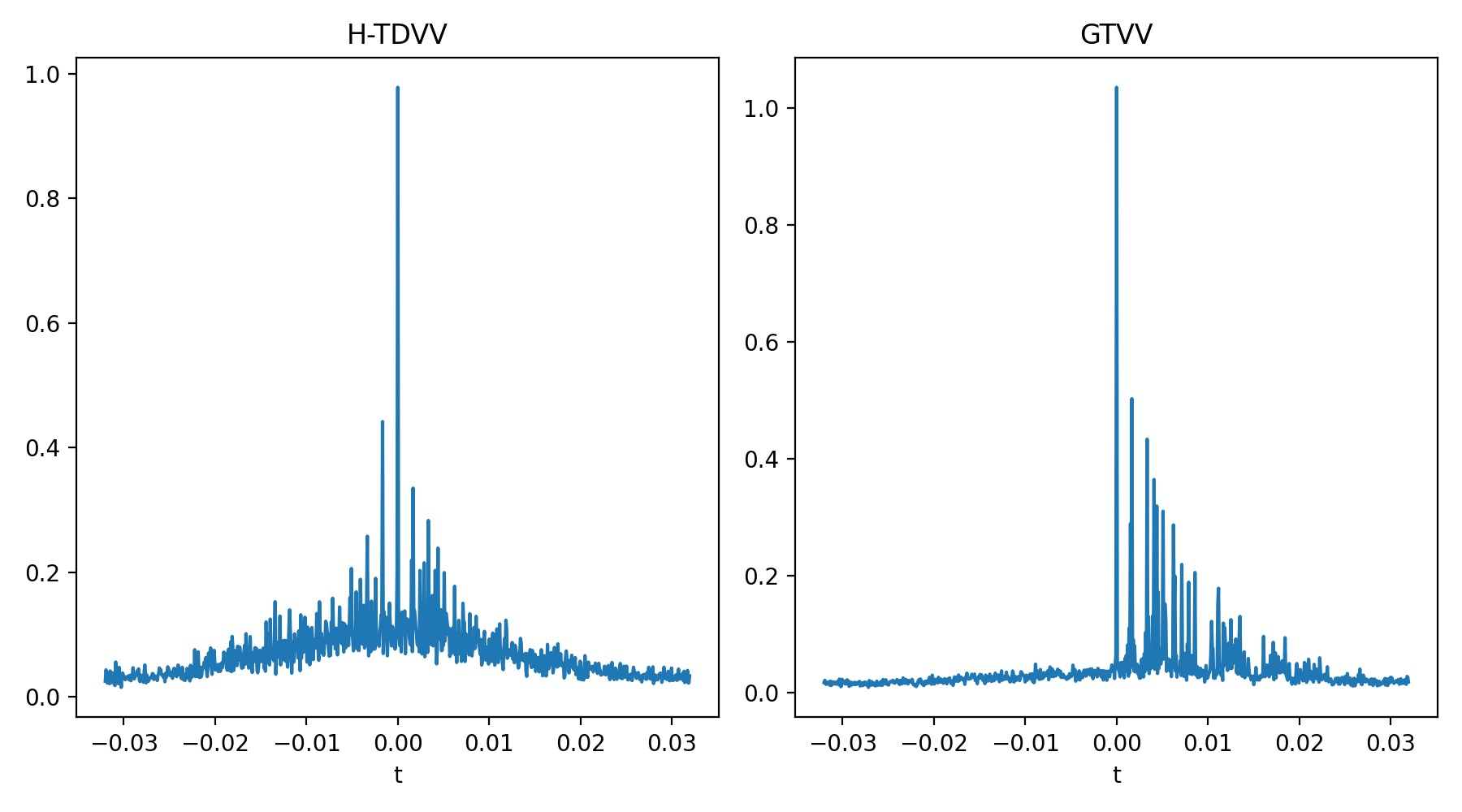}
    \caption{Magnitudes of H-TDVV (left) and GTVV (right) at different times $t$.}
    \label{figGTVV}
\end{figure}


Since $B^{(\ing{lm})}(\ing{u},f)$ and $U^{(\ing{lm})}(\ing{u},f)$ are correlated, $v^{(\ing{lm})}(f)$ and $U^{(\ing{lm})}(\ing{u},f)$ should be estimated simultaneously \cite{shalvi1996system}. Multiplying both sides of \eqref{eqRearranged} by $B^{(\ing{lm})}(f)^*$ and taking the expectation yields
\begin{multline}
    \phi_{B^{(\ing{lm})},B^{(\ing{lm})*}}(\ing{u},f) = \\
    v^{(\ing{lm})}(f)\sum\limits_{\ing{l}',\ing{m}'} w^{(\ing{l}'\ing{m}')}\phi_{B^{(\ing{l}'\ing{m}')},B^{(\ing{lm})*}}(\ing{u},f) + \phi_{U^{(\ing{lm})},B^{(\ing{lm})*}}(\ing{u},f),
\end{multline}
with $\phi_{XY} = \mathbb{E}[XY]$, and $\hat{\phi}_{XY}$ being its estimate. Assuming that noise statistics varies slower than that of speech, one can approximate that $\phi_{U^{(\ing{lm})},B^{(\ing{lm})*}}(\ing{u},f) \approx \phi_{U^{(\ing{lm})},B^{(\ing{lm})*}}(\ing{u}',f)$ holds within certain temporal segment $\ing{u}'\in[\ing{u}-\ing{u}_0,\ing{u}]$. This gives rise to the overdetermined system of equations
\begin{equation}\label{eqGVVestimate}
    \hat{\vect{\phi}}_{B^{(\ing{lm})},B^{(\ing{lm})*}} = \left[ \begin{matrix} \hat{\mtrx{\Phi}}_{B^{(\ing{lm})}} \vect{w} & \vect{1} \end{matrix} \right] \left[ \begin{matrix} v^{(\ing{lm})}(f) \\ \phi_{U^{(\ing{lm})},B^{(\ing{lm})*}}(f)  \end{matrix} \right] + \vect{\varepsilon},
\end{equation}
where $\hat{\vect{\phi}}_{B^{(\ing{lm})},B^{(\ing{lm})*}}$ and $\vect{\varepsilon}$ are the vectors of concatenated estimates of $\phi_{B^{(\ing{lm})},B^{(\ing{lm})*}}(\ing{u}',f)$, and errors $\varepsilon(\ing{u}',f)=\phi_{U^{(\ing{lm})},B^{(\ing{lm})*}}(f) - \hat{\phi}_{U^{(\ing{lm})},B^{(\ing{lm})*}}(\ing{u}',f)$, respectively, for ${\ing{u}'\in[\ing{u}-\ing{u}_0,\ing{u}]}$.  Likewise, $\hat{\mtrx{\Phi}}_{B^{(\ing{lm})}}$ is a matrix whose columns correspond to the estimates of cross-correlation between $B^{(\ing{lm})*}(\ing{u}',f)$ and $B^{(\ing{l}'\ing{m}')}(\ing{u}',f)$, for all $(\ing{l}',\ing{m}')$, and the same temporal segment, while $\vect{1}$ is the all-one vector. Solving such a system in the least-squares sense (with respect to $\vect{\varepsilon}$), for all $f$ and $(\ing{l},\ing{m})$, and applying the inverse Fourier transform, yields a robust estimate of $\vect{v}(t)$, $t\in(-T/2,T/2]$, where $T$ is the support size of the STFT window. Therefore, in practice, a GTVV estimate is a \emph{matrix} $\mtrx{V} \Rset{(L+1)^2}{T}$. 

\begin{algorithm}[t]
    \caption{S-OMP for the inference of GTVV parameters}\label{GOMP}
    \begin{algorithmic}[1]
    \State $\ing{i}=0, \; \mtrx{R}^{(0)} = \mtrx{V}, \; \hat{\mtrx{\Lambda}}=\{ \emptyset \}$
    \Repeat
    \State $\ing{i} \gets \ing{i}+1$
    \State $s^{(\ing{i})} = \argmax_{s} \|  \transp{{\mtrx{R}^{(\ing{i}-1)}}} \mtrx{Y}_{:,s} \|_{\infty}, \; \; \mtrx{Y}_{:,s} = \vect{y}(\Omega_{s}')$\label{sMaximize}
    \State $\hat{\mtrx{\Lambda}} \gets \hat{\mtrx{\Lambda}} \cup \Omega_{s^{(\ing{i})}}'$
    \State $\tau^{(\ing{i})} := \bar{\tau}^{(\ing{i})} - \bar{\tau}_0 = \frac{1}{f_s} \argmax_{\ing{q}} | \transp{\vect{y}(\Omega_{s^{(\ing{i})}}')}  \mtrx{R}_{:, \ing{q}}^{(\ing{i}-1)} |$\label{sTau}
    \State $\mtrx{Z}^{(\ing{i})} = \argmax_{\mtrx{Z}} \| \mtrx{Y}_{\hat{\mtrx{\Lambda}}} \mtrx{Z} -  \mtrx{V} \|_{\text{F}}^2$ 
    \label{sProject}
    \State $\mtrx{R}^{(\ing{i})} = \mtrx{Y}_{\hat{\mtrx{\Lambda}}}\mtrx{Z}^{(\ing{i})} - \mtrx{V}$\label{sResidual}
    \Until{stopping criterion}
    \State \textbf{return} $\hat{\mtrx{\Lambda}}, \{\bar{\tau}^{(\ing{i})} \}$
    \end{algorithmic}
\end{algorithm}

Examples of the estimated GTVVs for $\vect{w}=\transp{[1 \; \transp{\vect{0}}]}$ (termed here ``H-TDVV'' as an HOA extension of the TDVV model~\cite{daniel2020time}), and ${\vect{w}=\vect{y}(\Omega_0)}$ (the maximum directivity beamformer \cite{jarrett2017theory}, oriented toward DoA), are given in Fig.~\ref{figGTVV}.

\begin{table*}
    \centering
    \begin{tabular}{|r|c c c|c c c|c c c|c c c|}
    \hline
        HOA order & \multicolumn{3}{|c|}{1} & \multicolumn{3}{|c|}{2} & \multicolumn{3}{|c|}{3} & \multicolumn{3}{|c|}{4}  \\
        \hline \hline 
        \multicolumn{13}{|c|}{Low reverberation} \\
        \hline
        H-TDVV & $12.8^{\circ}$ & $0.17$ &  $2\cdot10^{-3}$ s & $15.0^{\circ}$ & $0.47$ & $5.1\cdot10^{-4}$ s & $10.9^{\circ}$ & $0.52$ & $10^{-4}$ s & $6.1^{\circ}$ & $0.87$ &  $\mathbf{6.1\cdot10^{-5}}$ s\\
        \hline
        GTVV &  $\mathbf{12.7}^{\circ}$ &  $\mathbf{0.20}$ & $\mathbf{10^{-4}}$ s & $\mathbf{9.8}^{\circ}$ & $\mathbf{1.01}$ & $\mathbf{9.9\cdot10^{-5}}$ s & $\mathbf{8.3}^{\circ}$ & $\mathbf{1.70}$ & $\mathbf{6.3\cdot10^{-5}}$ s & $\mathbf{5.8}^{\circ}$ & $\mathbf{3.12}$ & $6.7\cdot10^{-5}$ s \\
        \hline
        \multicolumn{13}{|c|}{High reverberation} \\
         \hline 
        H-TDVV & $14.6^{\circ}$ & $\mathbf{0.06}$ & $1.7\cdot10^{-2}$ s & $14.7^{\circ}$ & $0.18$ & $3.8\cdot10^{-3}$ s & $11.9^{\circ}$ & $0.22$ & $1.9\cdot10^{-4}$ s & $8.0^{\circ}$ & $0.12$ & $\mathbf{2.6\cdot10^{-5}}$ s \\
        \hline
        GTVV & $\mathbf{13.9}^{\circ}$ & $\mathbf{0.06}$ & $\mathbf{4.7\cdot10^{-4}}$ s & $\mathbf{12.6}^{\circ}$ & $\mathbf{0.77}$ & $\mathbf{3.3\cdot10^{-4}}$ s & $\mathbf{10.9}^{\circ}$ & $\mathbf{0.77}$ & $\mathbf{10^{-4}}$ s & $\mathbf{6.1}^{\circ}$ & $\mathbf{2.21}$ & $1.1\cdot10^{-4}$ s \\
        \hline
        
    \end{tabular}
    \caption{Reflection estimation performance (from left to right of each cell: angular error, number of detections and the delay error).}
    \label{tab:detection}
\end{table*}

\section{GTVV parameter inference via S-OMP}
\label{sec:inference}

Once the GTVV has been estimated, one can exploit its structure \eqref{eqGTVV} to obtain the acoustic parameters of individual wavefronts. For that purpose, we adopt the Simultaneous Orthogonal Matching Pursuit (S-OMP) algorithm \cite{tropp2006algorithms}, a greedy method for estimating the sparse support shared by a group of observation vectors. The aim of using such a simple algorithm is to emphasize the performance gains due to the presented GTVV model. Its pseudocode is given in Algorithm~\ref{GOMP}, which we describe and motivate in the following. 

Let $\mtrx{Y}\Rset{(L+1)^2}{Y}$ be a dictionary whose columns (``atoms'') are SH vectors\footnote{We use (without loss of generality) real-valued SH functions.} $\{ \vect{y}(\Omega_{\ing{j}}') \}_{\ing{j}\in[0,\ing{Y}-1]}$, obtained by sampling $\ing{Y}$ directions ${\mtrx{\Omega}' = \{ \Omega_j'\}}_{\ing{j}\in[0,\ing{Y}-1]}$ quasi-uniformly on the unit sphere. The model \eqref{eqGTVV} suggests that the GTVV matrix $\mtrx{V}$ could be approximated by a linear combination of few atoms from $\mtrx{Y}$, \emph{i.e.}, $\mtrx{V} \approx \mtrx{Y}_{\mtrx{\Lambda}}\mtrx{Z}$, where $\mtrx{Y}_{\mtrx{\Lambda}}$ is the set of columns of $\mtrx{Y}$ corresponding to the support $\mtrx{\Lambda}$, and $\mtrx{Z}\Rset{\#\mtrx{\Lambda}}{T}$ represents the contributions of different wavefronts. We assume that the number $\ing{N}$ of strong reflections in \eqref{eqMixtureModel} is small compared to the number of dictionary atoms $\ing{Y}$, \emph{i.e.} a sparse (synthesis) model. 

The orthogonalization step \ref{sProject} of Algorithm~\ref{GOMP} projects the observed $\mtrx{V}$ onto the subspace spanned by $\mtrx{Y}_{\mtrx{\Lambda}}$, where $\hat{\mtrx{\Lambda}}$ is the estimated support at the current iteration. The new residual $\mtrx{R}^{(\ing{i})}$ now contains only the contributions of wavefronts from directions $\mtrx{\Omega}'\setminus\hat{\mtrx{\Lambda}}$. In a favorable case, the selection rule in the step \ref{sMaximize}, based on the $\ell_{\infty}$ norm, would choose an atom $\vect{y}(\Omega_{s^{(\ing{i})}}')$ the most aligned with the wavefront $\vect{y}(\Omega_{\ing{n}})$, bearing the largest magnitude in this residual mixture. The pair $\Omega_{s^{(\ing{i})}}'=(\theta_{s^{(\ing{i})}}',\varphi_{s^{(\ing{i})}}')$ is then added to the set of estimated directions. However, should a scalar product with some of the ``cross-terms'' $\tilde{\eta}$ in \eqref{eqGTVV} become larger than this value, the selected atom will be erroneous. This becomes less likely for higher Ambisonic orders, due to the completeness property of spherical harmonics \cite{jarrett2017theory}.

According to \eqref{eqGTVV}, the GTVV magnitudes associated with a SH vector $\vect{y}(\Omega_{\ing{n}>0})$ are strictly decaying (due to ${|g_n \beta_n| < 1}$). Hence, the largest correlation between $\vect{y}(\Omega_{\ing{n}})$ and the current residual matrix $\mtrx{R}^{(\ing{i}-1)}$ should be observed for a column corresponding to $t\approx \tau_{\ing{n}} := \bar{\tau}_{\ing{n}} - \bar{\tau}_0$, which allows for the (relative) delay estimation in the step~\ref{sTau} (with $f_s$ being the sampling rate). Finally, under the assumption that the DoA component is the dominant wavefront, the dictionary atom chosen in the first iteration should coincide with $\Omega_0$ at $\bar{\tau}=0$.

\section{Computer experiments}
\label{sec:experiments}

\begin{table}[]
    \centering
    \begin{tabular}{|r|c|c|c|c|}
        \hline
        HOA order & 1 & 2 & 3 & 4  \\
        \hline \hline 
        \multicolumn{5}{|c|}{Low / High reverberation} \\
        \hline
        TRAMP & $18.8^{\circ}$/$29.7^{\circ}$ & $11.5^{\circ}$/$14.9^{\circ}$ & $10.9^{\circ}$/$14.2^{\circ}$ & $8.1^{\circ}$/$10.2^{\circ}$ \\
        \hline
        H-TDVV & $14.6^{\circ}$/$25.9^{\circ}$ & $4.5^{\circ}$/$6.2^{\circ}$ & $7.0^{\circ}$/$10.5^{\circ}$ & $3.0^{\circ}$/$5.7^{\circ}$ \\
        \hline
        GTVV & $\mathbf{7.5}^{\circ}$/$\mathbf{18.9}^{\circ}$ & $\mathbf{3.5}^{\circ}$/$\mathbf{5.3}^{\circ}$ & $\mathbf{3.7}^{\circ}$/$\mathbf{7.4}^{\circ}$ & $\mathbf{2.6}^{\circ}$/$\mathbf{4.8}^{\circ}$ \\
        \hline
    \end{tabular}
    \caption{DoA estimation error (best results are in bold).}
    \label{tab:direction}
\end{table}

To evaluate the presented GTVV model, we use the shoebox acoustic simulator software MCRoomSim  \cite{wabnitz2010room} to generate Room Impulse Responses (RIRs). This provides us with the complete ground truth data, containing not only the DoA information, but also the delays and directions of reflected components. The shoebox room dimensions are \mbox{$5\times4\times2.8$ m\textsuperscript{3}}, and the reverberation time is about $0.16$ s and $0.44$ s, for the low- and high-reverberant conditions, respectively. The received signals are obtained by convolving five RIRs, for each reverberation time condition, with a dry speech source at the sampling rate \mbox{$f_s=16$ kHz}. In order to simulate the diffuse interference, white Gaussian noise (in the Ambisonic encoding format) is added to each microphone channel, such that \mbox{SNR$=20$ dB}. The STFT frames (multiplied by the Hamming window) are $0.064$ long, and overlap by $75\%$.

The baselines are the HOA version of the TRAMP algorithm \cite{kitic2018tramp} (an efficient variant of the Ambisonic SRP-PHAT beamformer), and S-OMP applied to the ``omnidirectional'' \mbox{H-TDVV}. Moreover, we use the latter's DoA estimate $\hat{\Omega}_0$ to steer the beamformer ${\vect{w}=\vect{y}(\hat{\Omega}_0)}$ when computing GTVV. For simplicity, we consider only the DoA and first order reflections, despite the fact that \mbox{S-OMP} is oblivious to the reflection order. Thus, we choose the number of iterations (\emph{i.e.}, the number of estimated directions) as a simple stopping criterion. Clearly, S-OMP cannot estimate more directions than there are Ambisonic channels, hence we limit their number to $4$ (in the case of FOA), and $7$ (otherwise). 
All methods use the same dictionary $\mtrx{Y}$, obtained by discretizing $\ing{Y}=770$ directions on the spherical Lebedev grid \cite{lebedev1999quadrature}.

The DoA estimation performance, in terms of average angular error per frame, is presented in Table~\ref{tab:direction}. Likewise, the errors in estimating reflections are provided in Table~\ref{tab:detection}. Since the algorithm does not discriminate first order reflections, the results are restricted to those deviating no more than $20^{\circ}$ from the ground truth. Thus, we additionally provide the average detection results, as well as the delay estimation errors. 

The presented results indicate that all methods, generally, tend to perform better at higher Ambisonic orders, and worse when reverberation is significant (this is especially the case for FOA). Due to spatial limitations we are unable to show the results of the experiments with varying SNR levels. We report, though, that lowering SNR mainly impacts the detection rate, while the angular accuracy remains relatively stable.  Overall, the results confirm the predicted theoretical advantages of GTVV, as it largely outperforms the other two methods in all tested scenarios.

%
%
%
%
%
%

\section{Conclusion}
\label{sec:conclusion}


We have presented a Generalized Time Domain Velocity Vector, an extension of the TDVV representation introduced in the earlier work. GTVV is better conditioned than TDVV, and more potent at estimating individual wavefronts of the recorded Ambisonic mixture, as confirmed by computer experiments using the S-OMP algorithm. The future investigations will focus on more elaborate inference techniques, relaxation of some working assumptions, and the quantification of uncertainty of estimated acoustic parameters. 

\bibliographystyle{IEEEtran}
\bibliography{biblio}

\end{document}